\documentclass[aps,prl,twocolumn,superscriptaddress,floatfix,showpacs]{revtex4}

\usepackage{float}
\usepackage{placeins}
\usepackage{amssymb,amsmath}
\usepackage{graphicx}
\usepackage[colorlinks=true,linkcolor=blue,citecolor=blue,urlcolor=blue]{hyperref}

\begin{document}

\title{Current driven pinning strength in the vortex lattice of Nb$_3$Sn aided by a small oscillating magnetic field}

\author{M.\ Reibelt}
\email[]{reibelt@physik.uzh.ch}
\affiliation{Physik-Institut University of Zurich, Winterthurerstrasse 190, CH-8057 Zurich, Switzerland}
\author{N.\ Toyota}
\affiliation{Physics Department, Graduate School of Science, Tohoku University, 980-8571 Sendai, Japan}
\date{\today}

\begin{abstract}
By the application of a small oscillating magnetic field parallel to the main magnetic field and perpendicular to the transport current, we were able to generate a voltage dip in the I-V curves of Nb$_3$Sn similar to the peak-effect pattern observed in earlier resistivity measurements. The pattern was history dependent and exhibited a memory effect. In addition we observed in the I-V curves for a high shaking-field amplitude a step feature of unknown origin.
\end{abstract}


\maketitle

\begin{figure}
\includegraphics[width=75mm,totalheight=200mm,keepaspectratio]{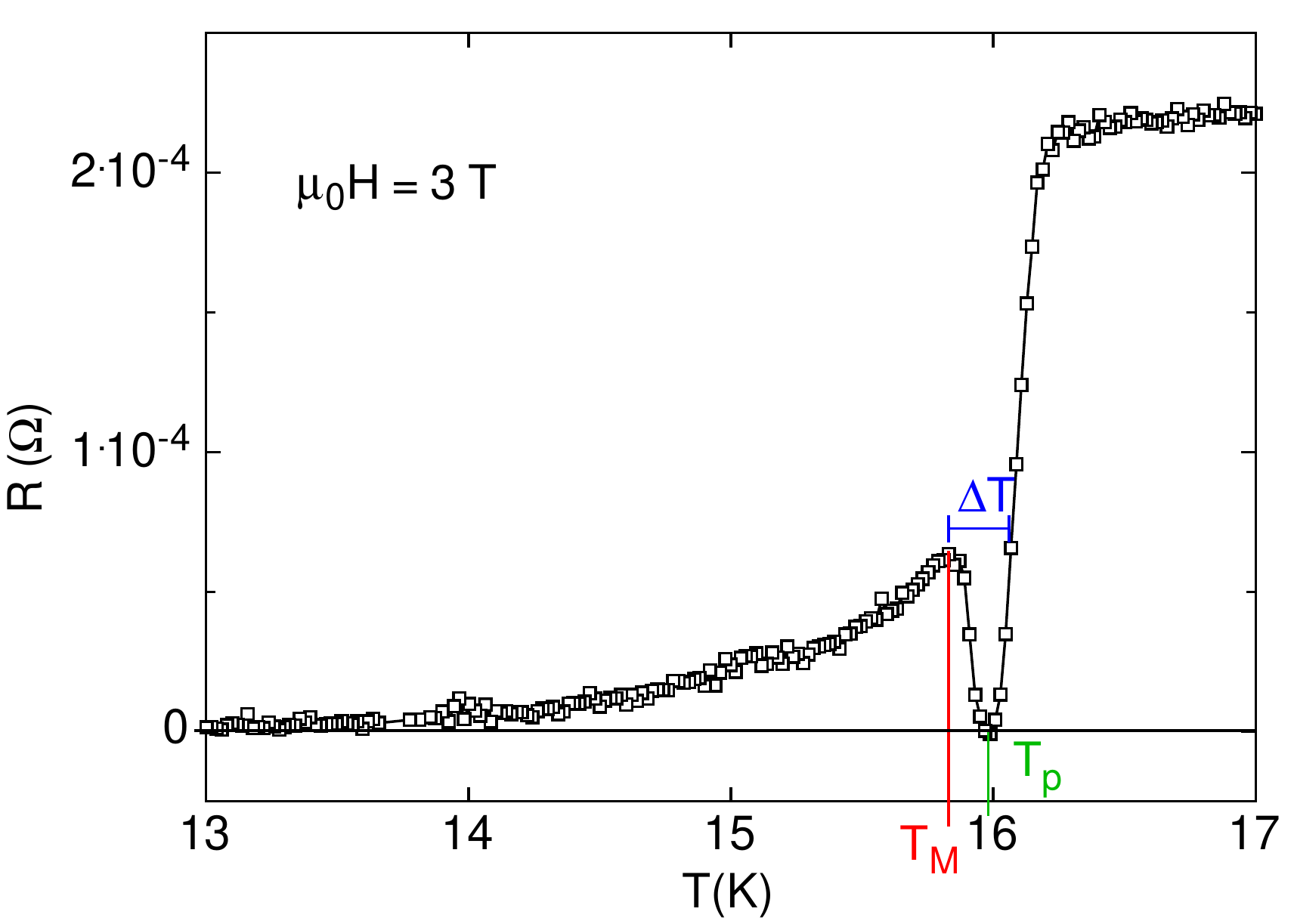}
\caption{Temperature dependence of the resistance $R(T)$ for a fixed external magnetic field $\mu_0 H = 3\, $T, transport current $I_m = 5\, $mA, and with with an oscillating magnetic field ($f = 1\, $kHz, $\mu_0 h_{ac} = 0.58\, $mT) superimposed. \label{fig.PRL2011_1_pdf}}
\end{figure}

\begin{figure}
\includegraphics[width=75mm,totalheight=200mm,keepaspectratio]{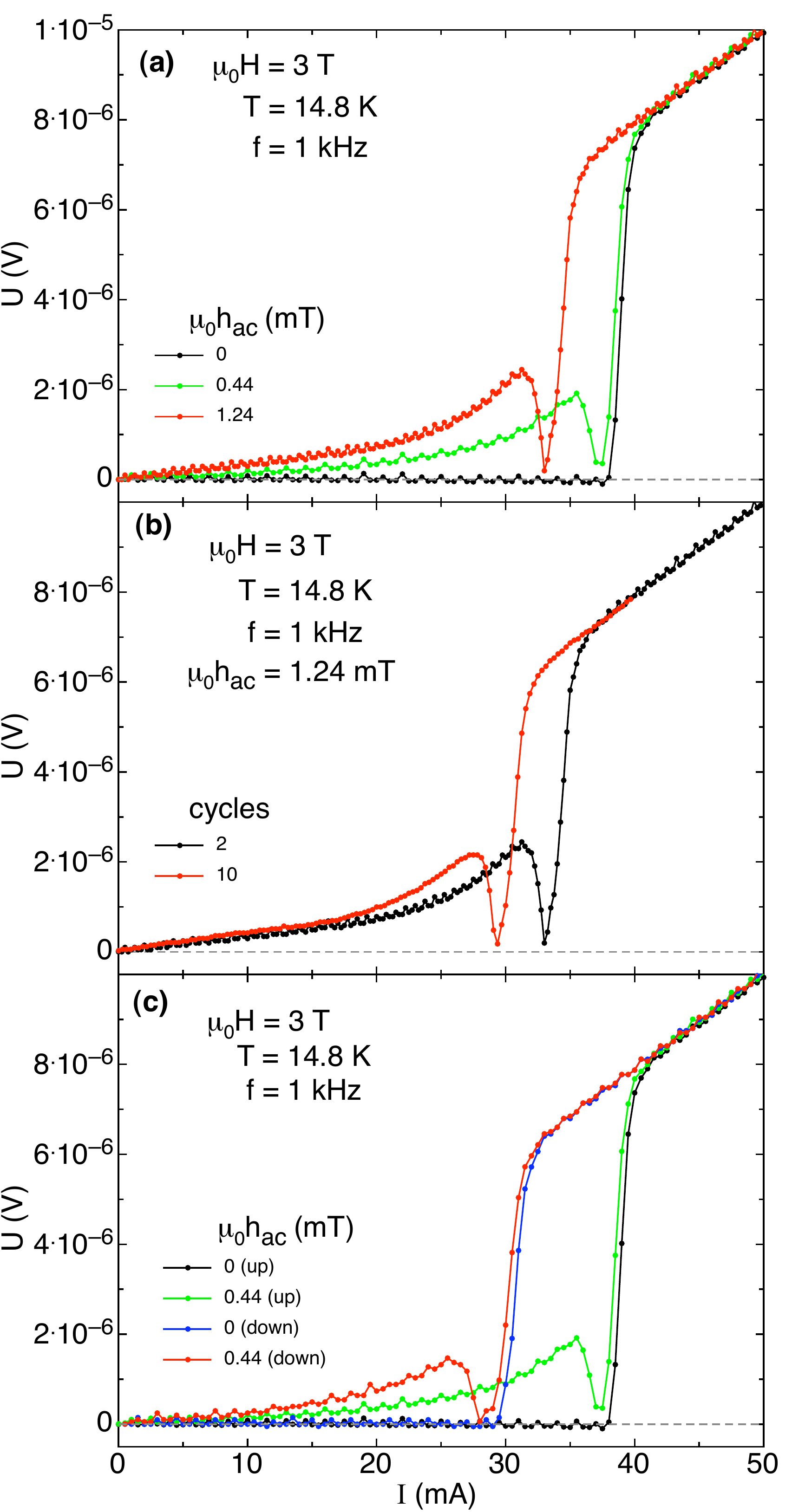}
\caption{I-V curves at fixed temperature $T=14.8\, $K and magnetic field $\mu_0H=3\, $T. \textbf{(a)} The shaking-field amplitude $h_{ac}$ is varied (fixed frequency $f=1\, $kHz). \textbf{(b)} The number of power line cycles during the averaging process for each data point is varied. \textbf{(c)} I-V curves for increasing and decreasing current. \label{fig.PRL2011_2_pdf}}
\end{figure}

\begin{figure}
\includegraphics[width=75mm,totalheight=200mm,keepaspectratio]{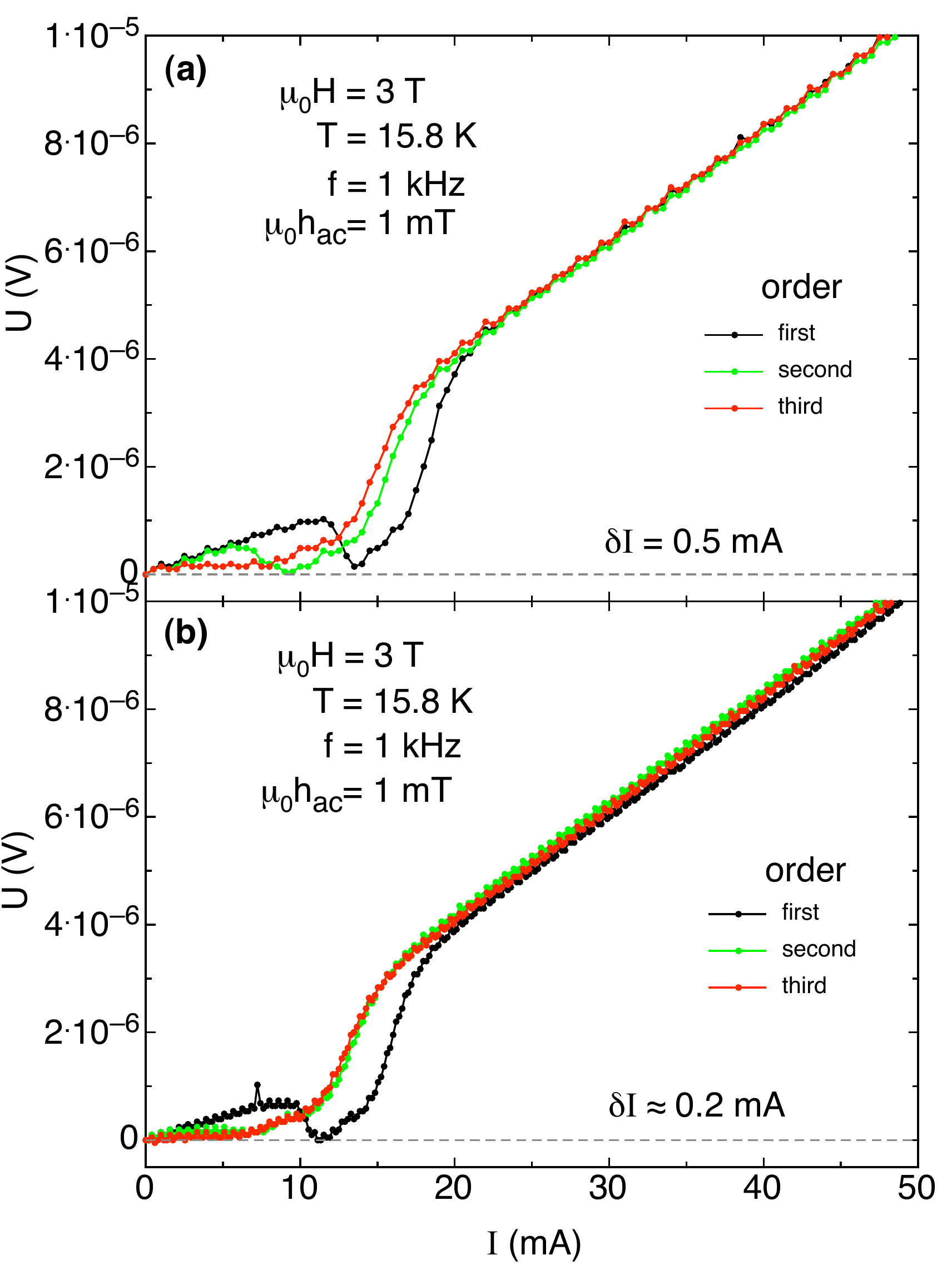}
\caption{Subsequently recorded I-V curves at fixed temperature, main magnetic field, and shaking field; with current step-width \textbf{(a)} $\delta I = 0.5\, $mA, \textbf{(b)} $\delta I \approx 0.2\, $mA. \label{fig.PRL2011_4_pdf}}
\end{figure}

\begin{figure}
\includegraphics[width=75mm,totalheight=200mm,keepaspectratio]{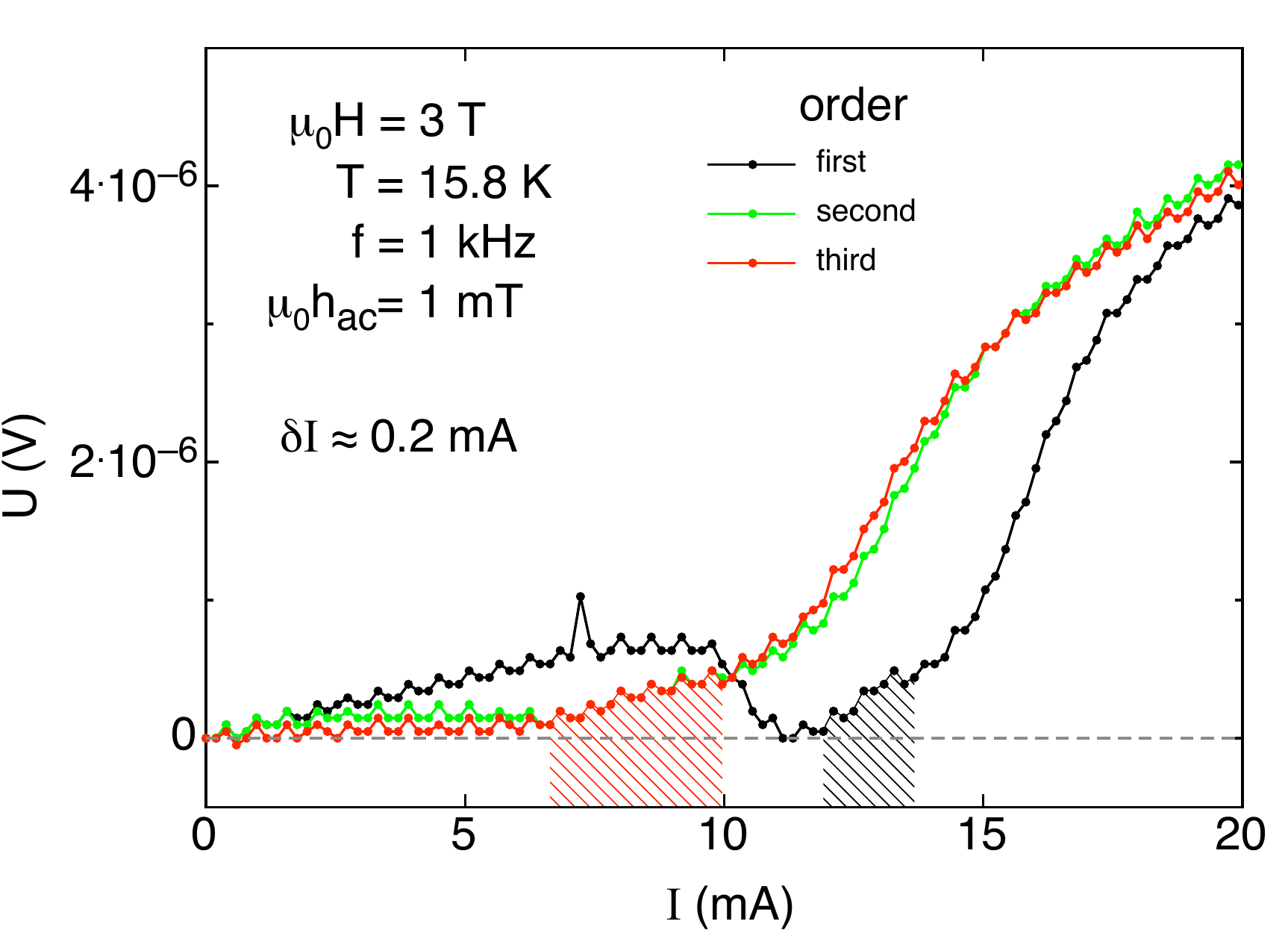}
\caption{Magnification of Fig.~\ref{fig.PRL2011_4_pdf}(b) at the peak effect region. The step feature is marked by dashed areas. \label{fig.PRL2011_5_pdf}}
\end{figure}

The vortex dynamics near the peak effect in type-II superconductors is a field of ongoing research since many years \cite{Higgins1996}. The peak effect is the term used for the sudden increase of the critical current due to increased pinning in a small region of the magnetic phase diagram just below the upper critical field $H_{c2}$. Besides in ac susceptibility \cite{Marchevsky2001February,Adesso2006March} and dc magnetization \cite{Lortz2007March}, the peak effect has also been observed in electrical-transport measurements \cite{Huebener1970February,Meier-Hirmer1985January,Wordenweber1986March,Higgins1996,Henderson1996September,Sato1997,Langan1997,Chaudhary2001,Daniilidis2007}. In most transport measurements either the temperature or the magnetic field was varied; however, there are some experiments in which the transport current was varied. While Uksusman \emph{et al.}~\cite{Uksusman2009} did not observe a peak-effect pattern in their I-V curves in YBa$_2$Cu$_3$O$_{7-\delta}$ tapes, Marchevsky \emph{et al.}~\cite{Marchevsky2002February} were able to observe a counter-intuitive increase of pinning strength on increasing the transport current in a thin strip of NbSe$_2$. Marchevsky \emph{et al.}~\cite{Marchevsky2001February,Marchevsky2002February} described their data and the peak effect with a two-phase model in line with the edge-contamination mechanism proposed by Paltiel \emph{et al.}~\cite{Paltiel2000January,Paltiel2000October}, where a high-pinning phase enters the sample on increasing the current from the edges, thereby causing an inhomogeneous current distribution inside the sample. In addition, an anomalous pinning enhancement during the superconducting-normal transition of $2$H-NbSe$_2$ was observed by Yue \emph{et al.}~\cite{Yue2004}.\\
\indent According to theories \cite{Brandt1986November,Brandt1994April,Brandt1996March,Brandt1996August,Mikitik2000September,Mikitik2001August,Brandt2002July,Brandt2004,Brandt2004a,Brandt2004b,Brandt2007}, transversal and also longitudinal vortex shaking with a small oscillating magnetic field (so-called shaking field) $h_{ac}$ can cause magnetic vortices to "walk" through the superconductor, which leads to a reduction of nonequilibrium current distributions and thereby to an annealing of the vortex lattice. This technique was successfully applied, for example, for resistivity measurements \cite{Cape1968February}, torque magnetometry \cite{Willemin1998September,Weyeneth2009}, local magnetization measurements \cite{Avraham2001May}, and also for specific-heat investigations \cite{Lortz2006September}. In this work, we measured I-V curves on a Nb$_3$Sn single crystal while simultaneously applying a shaking field $h_{ac}$. A configuration that is comparable to ours and with a similar shaking technique was used in \cite{Huebener1971,Risse1997June,Uksusman2009}. We report on the appearance of a peak-effect pattern in these I-V curves.\\
\indent The Nb$_3$Sn single crystal used for this study (mass $m \approx 11.9\, $mg, thickness $d \approx 0.4\, $mm, and cross section $A \approx 0.44\, $mm$^2$) was characterized by Toyota \emph{et al.}~\cite{Toyota1988September} and was further investigated by Lortz \emph{et al.}~for calorimetric experiments \cite{Lortz2006September,Lortz2007April,Lortz2007March}. The crystal exhibits a pronounced peak effect near $H_{c2}$ \cite{Lortz2007March}, and its transition to superconductivity in zero magnetic field as determined by a resistivity measurement occurs at $T_{c} \approx 18.0\, $K.\\
\indent The I-V curves were measured in a four-point configuration in a commercial physical property measurement system (PPMS) from Quantum Design. The data were collected during increasing the current in small steps from zero to $60\, $mA or on decreasing the current to zero. The current step-width varied between $0.25\, $mA and $0.8\, $mA. The setup was the same as in \cite{Reibelt2010March}. The external main magnetic field was applied perpendicular to the flat side of the crystal $(H \parallel c)$ and also perpendicular to the transport current. We mounted an ac shaking coil with $1306$ windings on a commercial PPMS resistivity puck with its axis aligned with the main magnetic field so that $h_{ac} \parallel H$. A theoretical treatment of this configuration was done by Mikitik \emph{et al.}~\cite{Mikitik2001August}. In order to monitor the amplitude of the shaking field $h_{ac}$ we installed an additional pick-up coil inside this ac shaking coil. To achieve an accurate measurement of the sample temperature, we attached a calibrated thermometer (Cernox, LakeShore Inc.) at the top of a brass block supporting the sample. The ac shaking coil was driven by an external ac power source (Agilent model $6811$B).\\
\indent In order to give the reader an orientation for the in the following presented measurements, we have replotted in Fig.~\ref{fig.PRL2011_1_pdf} the temperature sweep for a resistivity measurement at fixed main magnetic field $\mu_0 H=3\, $T which was already shown in Fig.~$9$ of \cite{Reibelt2010March}, where $T_p$ is the temperature of the dip which coincides with the peak effect, $\Delta T$ is the width of this dip and $T_M$ is the temperature where the resistivity reaches a local maximum, which is also the onset temperature of the peak-effect dip.\\
\indent In Fig.~\ref{fig.PRL2011_2_pdf}(a) we plotted I-V curves at fixed temperature $T = 14.8\, $K and fixed main magnetic field $\mu_0 H=3\, $T; the shaking-field amplitude $\mu_0h_{ac}$ varied between zero and $1.24\, $mT while the frequency of the shaking field was held constant at $f = 1\, $kHz. The critical current for the measurement without shaking field is about $38\, $mA (black curve). For the I-V curves with simultaneously applied shaking field, a distinct voltage-dip pattern emerges which resembles very much the peak-effect pattern which we have observed before in resistivity measurements \cite{Reibelt2010March} (see Fig.~\ref{fig.PRL2011_1_pdf}). For low currents the resistive voltage increases linearly with the current, followed by a region where the voltage increases stronger than linear before it reaches a maximum, after which a sharp dip follows where the voltage returns to nearly zero or eventually even zero. After the dip the vortex lattice subsequently enters the flux-flow region where the I-V curve has a constant slope, the flux-flow resistance is $\rho_f \approx 2.1*10^{-4}\, $VA$^{-1}$, which is as expected nearly as large as the normal resistance $\rho_n$ in Fig.~\ref{fig.PRL2011_1_pdf}. All I-V curves match in the flux-flow region. Increasing $h_{ac}$ from $0.44$ to $1.24\, $mT shifts the I-V curve to lower currents and increases its slope in the linear region at low currents. These observations may be interpreted as follows. Already at low currents a fraction of the vortex lattice is depinned by the combination of the transport current and the shaking field, which causes the vortices to "walk" across the sample, which in turn leads to a finite resistance which manifests as a constant slope in the I-V curve. According to Brandt and Mikitik \cite{Brandt2002July}, the by the "walking" of the vortices generated electric field and therefore the resistance is proportional to $h_{ac}$, in line with our observation, that the slope of the I-V curves increases with $h_{ac}$ in their linear part for low currents (compare I-V curves in Fig.~\ref{fig.PRL2011_2_pdf}(a)). At about $16\, $mA the combined action of current and shaking field starts to further anneal the vortex lattice which leads to an increase in flux motion across the sample and thereby increases its resistance which manifests in the I-V curves as a strong nonlinear increase of the voltage until a maximum is reached.
At this maximum, further increase of the current leads to a counterintuitive decrease in the voltage which drops nearly to zero or perhaps even to zero due to an increase in vortex pinning. A similar counterintuitive current induced pinning increase of the vortex lattice was observed by Marchevsky \emph{et al.}~\cite{Marchevsky2002February}, who interpreted this effect as the consequence of a current driven injection of a high-pinning phase entering from the edges. However, their observed increase in pinning strength is less pronounced than in our experiments (see Fig.~$2$(b) in \cite{Marchevsky2002February}), which may be due to the circumstance that in their experiment only a small part of the sample was influenced by a shaking field.\\
\indent Usually the measurement of one data point took place by averaging over the period of two power line cycles. We replotted the I-V curve for $\mu_0h_{ac}=1.24\, $mT in Fig.~\ref{fig.PRL2011_2_pdf}(b) along with an additional I-V curve for which ten cycles per data point were averaged. Both curves start to deviate from linearity at about $16\, $mA, however, the ten-cycles I-V curve has a stronger slope increase in its nonlinear part, while both curves match for lower currents. This observation may support the interpretation that at about $16\, $mA an annealing process sets in which is time dependent and therefore has been advanced further for the ten-cycles I-V curve, leading to a higher voltage above $16\, $mA compared to the two-cycles I-V curve.\\
\indent In Fig.~\ref{fig.PRL2011_2_pdf}(c) we plotted data at $T=14.8\, $K and $\mu_0H=3\, $T which we collected on increasing and on decreasing the current. The down-sweep I-V curves are shifted to lower currents compared to the up-sweep I-V curves. This hysteresis may be explained with an analogy from mechanical friction. In order to set a body into motion against the mechanical friction, a larger force is needed than what is needed in order to sustain the motion of an already moving body. A similar mechanism may take place with the moving vortex lattice. For an already moving vortex lattice, the vortices cannot adjust perfectly to the pinning sites during the movement and therefore a smaller current I < Ic is able to sustain the vortex motion for the decreasing current sweep. Additional application of a shaking field also unveils the voltage-dip pattern in the down-sweep I-V curve, although the width of the dip appears to be broadened. This puzzling observation may be explained with the two vortex-phase mixture model which was also applied by Marchevsky \emph{et al.}~\cite{Marchevsky2002February}. In this model the vortex lattice gets "contaminated" by a high-pinning phase from the edge when an increasing driving current is applied thereby substituting the low-pinning phase. For a decreasing current the opposite process may take place; a low-pinning phase may substitute the high-pinning phase with decreasing current. This low-pinning phase in turn is more susceptible to the a.c. field which leads to the linear I-V-curve at low currents in the decreasing current sweep. However, this interpretation is speculative and should be further investigated in future experiments.\\
\noindent We want to stress that our procedure was able to generate a peak-effect pattern at the temperature $T=14.8\, $K, which is considerably lower than the temperature where the peak effect is usually situated for $\mu_0H=3\, $T ($\approx 16\, $K) (see Fig.~\ref{fig.PRL2011_1_pdf}). This finding supports the assumption of Marchevsky \emph{et al.}~\cite{Marchevsky2001February,Marchevsky2002February}, that the peak effect is a disorder-driven non-thermal phase transition.\\
\begin{figure}
\includegraphics[width=75mm,totalheight=200mm,keepaspectratio]{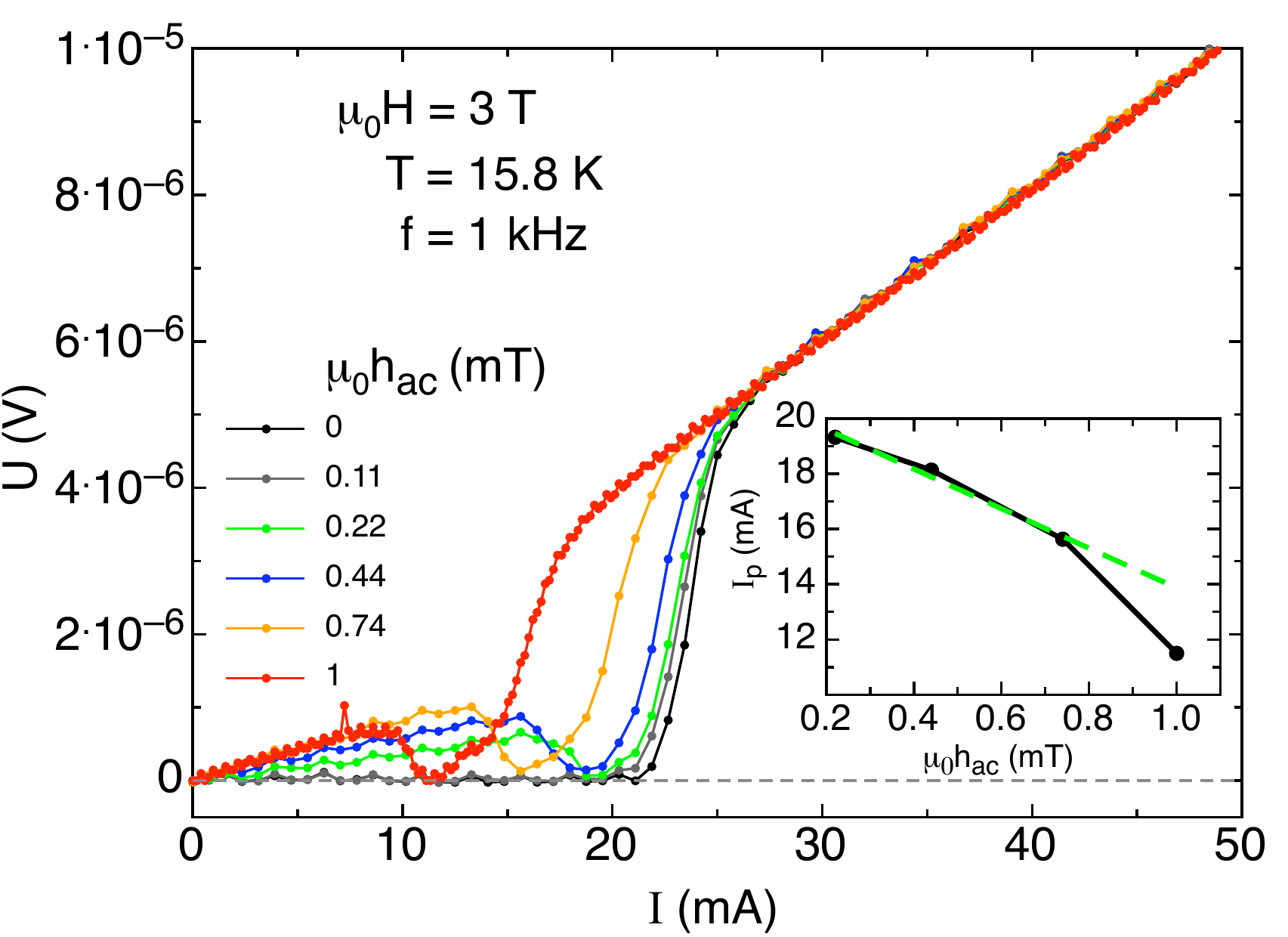}
\caption{I-V curves at fixed temperature $T=15.8\, $K and magnetic field $\mu_0H=3\, $T. The shaking-field amplitude $h_{ac}$ is varied (fixed frequency $f=1\, $kHz). The inset shows the linear dependence of $I_p$ on $h_{ac}$ (see text). \label{fig.PRL2011_3_pdf}}
\end{figure}
\indent Next, we shifted the temperature $1\, $K closer towards $H_{c2}(T)$. In Fig.~\ref{fig.PRL2011_3_pdf} we plotted I-V curves at $T=15.8\, $K and $\mu_0H=3\, $T for different shaking-field amplitudes. The critical current for the measurement without shaking field is about $22\, $mA (black curve). For the I-V curves with simultaneously applied shaking field, again a distinct voltage-dip pattern emerges, however, the I-V curves are shifted to lower currents compared to the I-V curves at $T=14.8\, $K and the nonlinear part is missing. The small shaking-field amplitude $\mu_0h_{ac}=0.11\, $mT (gray curve) is too small to create the voltage-dip pattern in the I-V curve, no resistive voltage below the critical current is detected, the I-V curve is merely shifted by about $1\, $mA to lower currents. The higher shaking-field amplitude $\mu_0h_{ac}=0.22\, $mT (green curve) then creates a voltage-dip pattern in the I-V curve, with a linear part for low currents below the maximum. This observation indicates that a threshold shaking-field amplitude exists between $0.11\, $mT and $0.22\, $mT in line with the threshold shaking-field amplitude we observed in \cite{Reibelt2010March}. With increasing $h_{ac}$, the I-V curves are shifted to lower currents and the slope of the linear part increases, except for the I-V curve with $\mu_0h_{ac}=1\, $mT which has approximately the same slope in its linear part as the I-V curve for $\mu_0h_{ac}=0.74\, $mT. The inset in Fig.~\ref{fig.PRL2011_3_pdf} shows the linear dependence of the current $I_p$, which marks the position of the voltage dip, on the shaking-field amplitude $h_{ac}$. The I-V curve for $\mu_0h_{ac}=1\, $mT may deviate a bit from this linearity because it has a lower current step-width ($\delta I \approx 0.2\, $mA) than the other I-V curves ($\delta I \approx 0.8\, $mA) in Fig.~\ref{fig.PRL2011_3_pdf}. Strikingly, at about $13\, $mA a step feature occurs in the I-V curve for $\mu_0h_{ac}=1\, $mT (red curve). This step feature was reproducible (see Figs.~\ref{fig.PRL2011_4_pdf} and \ref{fig.PRL2011_5_pdf}), its origin is unknown, however. Although speculative, the position of the step feature inside the peak effect region may hint to a connection to the step change in the equilibrium diamagnetism due to a first-order phase transition, which is suspected to underlie the peak effect \cite{Zeldov1995June,Ravikumar2000December}. Whether this step feature is intrinsic and maybe even generic or due to an inhomogeneity of the sample should be investigated in the future by measurements on different samples of Nb$_3$Sn and other type-II superconductors.\\
\indent At last we investigated the history dependence of the I-V curves. In Fig.~\ref{fig.PRL2011_4_pdf}(a) we plotted three I-V curves with current step-width ($\delta I = 0.5\, $mA) which were recorded subsequently with a gap of about $20$ seconds between each of them. The I-V curves of the later runs are shifted to lower currents and the width of the voltage drop increases until for the last I-V curve there is no linear-increase region anymore and the voltage stays close to zero until the onset of the step feature close to the flux-flow region. The step feature occurs in all three I-V curves. Although the step feature is also shifted to lower currents for the later runs, its shape and size seems to stay uninfluenced. Fig.~\ref{fig.PRL2011_4_pdf}(b) shows the same kind of experiment but with a lower current step-width ($\delta I \approx 0.2\, $mA), which brought out the step feature even further. The interpretation of this history dependence may be as follows. The current reached in the flux-flow region does not completely anneal the high-pinning phase of the vortex lattice which was induced in the sample during the voltage-dip region. Some vortices or bundles of vortices may have stayed attached to strong pinning sites and did not participate in the flux flow. When the next run starts these bundles of vortices are already pinned at the strongest pinning sites and they support the formation of a new high-pinning phase as an initial nucleus, which is why the voltage dip broadens and sets in already for lower currents. However, this memory-effect interpretation is speculative.
Nevertheless, one can state that the susceptibility of the investigated Nb$_3$Sn sample to small oscillating magnetic fields at low currents can be reduced by this procedure of repeated current ramping in an a.c.~field, which may be of practical relevance for applications.\\
\\
\indent To conclude, we have reported on the, to the best of our knowledge, first observation of a peak-effect pattern in the I-V curves of Nb$_3$Sn, which was generated by the combined application of a transport current and a small shaking field. No corresponding pattern appeared in the I-V curves that were taken without such a shaking field. We investigated the effect at different temperatures and observed a memory effect. Our findings strengthen the assumption that the peak effect is a disorder-driven non-thermal phase transition. In addition we observed for a high shaking-field amplitude $h_{ac}=1\, $mT a reproducible step feature of unknown origin in the I-V curves that may be related to a first-order phase transition, which is suspected to underlie the peak effect.\\
\indent We thank the group of Prof.~Schilling for support. This work was supported by the Schweizerische Nationalfonds zur F\"{o}rderung der Wissenschaftlichen Forschung, Grants No.~$20$-$111653$ and No.~$20$-$119793$.


\end{document}